\documentstyle[12pt]{article}
\global\arraycolsep=2pt
 \newcommand{\be}{\begin{eqnarray}}
\newcommand{\ee}{\end{eqnarray}}

\begin{document}
  \begin{titlepage}

\begin{center}

{\Large \bf  Large Order Behavior of Quasiclassical Euclidean Gravity in Minisuperspace Models}
 \vspace{2cm}

  {\bf T. Fugleberg} and {\bf   A. Zhitnitsky\footnote{ also:
Budker Institute of Nuclear Physics, Novosibirsk, Russia, 63090}}
 \vspace{0.5cm}

 {\it Department of Physics and Astronomy, University of British Columbia, 6224 Agricultural Road, Vancouver, BC, V6T 1Z1,  Canada}
\date{\today}
\end{center}
\begin{abstract}
We demonstrate in two minisuperspace models that a perturbation expansion of
quasiclassical Euclidean gravity has a factorial dependence on the order of the term at
large orders.  This behavior indicates that the expansion is an asymptotic series which
is suggestive of an effective field theory.  The series may or may not be Borel summable
depending on the classical solution expanded around.  We assume that only the
positive action classical solution contributes to path integrals.
We close with some speculative discussion on possible implications of the
asymptotic nature of the expansion.\end{abstract}
\end{titlepage}
\vskip 0.3cm
\noindent

\section{Introduction}
The starting point of our analysis is the Euclidean action
of gravity \cite{Hawking1}:
\begin{equation}
\hat{I}=-\frac{1}{16\pi G}\int_{\cal{M}} (R-2\Lambda)\sqrt{g} d^4x. \end{equation}
This action comes from the Euclidean continuation of the
Hilbert action.
We include the cosmological constant term but we will be interested
mainly in $\Lambda$ near zero.
The Euclidean partition function is then:
\begin{equation}
Z=\int D[g] \exp (-\hat{I}[g]).
\label{partitionfunction}
\end{equation}
There are many problems with this definition
of gravity as a fundamental theory, (see \cite{GHPerry}),
which we are not going to discuss in this letter.  Rather, we want to look at gravity
as an effective field theory described by an effective Lagrangian, $L_{eff}$,
which, by definition, contains operators of arbitrary dimensionality.
This view of gravity as an effective field theory of an unknown high
energy fundamental theory
was advocated for a many years by Weinberg\cite{Weinberg}, see also the recent
paper\cite{Donoghue}. As was argued recently
\cite{Zhitnitsky}, the corresponding expansion in general
is not a convergent but rather an asymptotic series with factorially growing coefficients.
Let us note that this remark about the factorial dependence
of the series is an absolutely irrelevant issue for the
analysis of low energy phenomena.
We have nothing new to say about these issues. However, if we wish
to discuss Plank scale phenomena, we need to know the behavior of the whole series
when distant terms in the series
might be important. Only in this case does the analysis of the large
order terms in the expansion have some physical meaning.

The first application of this idea to problems of inflation
was discussed recently \cite{Brandenberger&Zhitnitsky}.  It was argued
that the effective potential which is obtained by resumming the series provides
a natural realization of the inflationary scenario.  Therefore, it would be
worthwhile to prove the main assumption of the paper \cite{Brandenberger&Zhitnitsky}
about factorial growth of coefficients in the effective lagrangian.
The {\bf main goal} of the present letter is
to explicitly demonstrate the factorial dependence of quasiclasssical Euclidean
gravity in two minisuperspace models.

The series may or may not be Borel summable depending on which classical
solution the path integral is expanded about.  We assume following
\cite{Marolf} that only classical solutions with positive action are
relevant.  It should be noted, however, that the factorial
dependence is independent of these details.
We discuss some applications in the conclusion.

\section{A Simple Minisuperspace Model}

Assume a metric of the form:
\begin{equation}
ds^2 = dt^2  +  a(t)^2 d\Omega^2.
\end{equation}
The Euclidean action in this case is:
\begin{equation}
\hat{I}=-\frac{3\pi}{4G}\int_0^T dt \sqrt{a^6} \left[ a^{-2} (1-\dot{a}^2) - \ddot{a}a^{-1} - \frac{\Lambda}{3} \right].
\label{action1}
\end{equation}
We consider this as an analytical functional of a complex scale factor
`a(t)'.  We stress this point since it is
usual in general relativity to consider only real scale factors, in which
case the action would be strictly negative after choosing the positive sign volume
form.  Treating `a(t)' as a complex number, however, we can write
the action in the form:
\begin{equation}
\label{qm}
\hat{I}=-\frac{3\pi}{4G}\int_0^T dt \left[ a (1-\dot{a}^2) - \ddot{a}a^2 - \frac{\Lambda}{3} a^3\right],
\label{action2}
\end{equation}
which can have arbitrary phase.  This agrees
with the action in \cite{KlebanovEtal} after integration by parts.
Both forms lead to the same results in the leading classical term.
At higher orders, surface terms which are neglected
in \cite{KlebanovEtal} become important.
Both forms of the action have vanishing variation for solutions to:
\begin{equation}
1- \dot{a}^2 - 2 a \ddot{a} - \Lambda a^2 = 0,
\end{equation}
and resulting classical solutions are $a_{cl}=\pm \sqrt{\frac{3}{\Lambda}} \sin(\sqrt{\frac{\Lambda}{3}}t)$.  The positive solution is the usual solution and
the only solution considered in \cite{KlebanovEtal}.  It gives a negative action.  The second solution produces a positive action.

In either case, restricting ourselves to a compact Euclidean geometry in the classical limit
by selecting the final time $T=\pi\sqrt{3/\Lambda}$, we obtain a universe that expands from zero size at t=0 to
a maximum and recontracts to a point at t=T.  The manifold described
by this metric is $S^4$ and the spacetime it describes is called Euclidean De Sitter space.

The surface terms from the previously mentioned integration
by parts vanish because $a_{cl}$ vanishes at the endpoints of the
integration.  Note also that the manifold has no boundary which means that
a boundary integration over extrinsic curvature (K), which we have neglected,
cannot contribute.

At this point we can substitute the value of the classical action into the path integral (\ref{partitionfunction}) to obtain in the classical approximation:
\begin{equation}
Z \propto \exp \left(\pm \frac{3\pi}{\Lambda G}\right)
\end{equation}
where the signs agree with the sign of the classical solution.
The existence of a classical solution with finite Euclidean action means that
this solution can be used to calculate the large order behavior of the
perturbation theory as was suggested many years ago \cite{Lipatov},\cite{Zinn-Justin&etal}.

Notice that as $\Lambda \rightarrow 0$ the first solution
blows up while the second is perfectly well behaved.  Klebanov et al. \cite{KlebanovEtal} used the first solution.  Since we are looking for a theory
that has meaning for vanishing cosmological constant, we choose to do a
quasiclassical approximation about the other classical solution, $a_{cl}= - \sqrt{\frac{3}{\Lambda}} \sin(\sqrt{\frac{\Lambda}{3}}t)$.  This is the same solution one would get by performing a Hawking
rotation on `a(t)' in the action integral, solving and then
continuing back to real `a(t)'(see \cite{KlebanovEtal}).  Thus, we obtain
the same result without using the Hawking prescription.

With only the classical solution we can now derive the factorial dependence
of the higher order terms in the quasiclassical expansion.  We would
like to consider the path integral as a function of a small coupling constant
$g=\Lambda/3$ and find its expansion in `g':
\begin{equation}
\label{k}
Z(g)=\int D[a]\, \exp \left( \frac{3\pi}{4G}\int_0^T dt \left[ a (1+\dot{a}^2) - g a^3\right]\right)\equiv \sum_{K=0}^\infty Z_K g^K.
\end{equation}
For large values of the classical action, this functional integral over complex
`a(t)' is dominated by the value of the integrand at the corresponding classical
solution.
Using the nontrivial value of the classical action and techniques from
\cite{Lipatov},\cite{Zinn-Justin&etal} (see also review \cite{Zinn-Justin}) we obtain:
\begin{equation}
Z_K\propto \int_{-\infty}^0\frac{dg}{g^{K+1}}\, \exp \left(\frac{\pi}{gG}
\right) =
(-\frac{G}{\pi})^K \Gamma(K) \approx (-\frac{G}{\pi})^K K!
\end{equation}
for large K when we expand about the negative action classical solution.  Therefore, this expansion has the promised factorial dependence.

Expanding about the positive action classical solution requires a modification
of the formulas from \cite{Zinn-Justin} but gives the following result for 
large K:
\begin{equation}
\label{1}
Z_K\propto \int_0^\infty\frac{dg}{g^{K+1}} \exp \left(-\frac{\pi}{gG}
\right) =
(\frac{G}{\pi})^K \Gamma(K) \approx (\frac{G}{\pi})^K K!.
\end{equation}

Notice that both cases have the factorial dependence characteristic of an asymptotic series.  The only difference between the two cases is that the
negative action solution has alternating sign whereas the positive action
solution does not.  The alternating sign could make the series Borel summable,
but we believe the classical solution with positive action is the physically
favorable solution\cite{Marolf}.  In principle all saddle points may contribute
to the functional integral and which saddle points actually do contribute is
determined by the definition of the functional integral.

The quasiclassical expansion about this solution is:
\begin{equation}
\hat{I}[a_{cl}+\delta a] \approx \hat{I}_{cl} + \frac{1}{2} \int_0^T dt \delta a(t)
\left[-2 \Lambda a_{cl} - 2\ddot{a}_{cl} - 2\dot{a}_{cl} \frac{d}{dt} - 2 a_{cl} \frac{d^2}{dt^2}\right] \delta a(t)
\label{expansion}
\end{equation}
The effect of quadratic fluctuations can be determined by expanding the
perturbation $\delta a(t)$ in an orthogonal set of eigenfunctions
of the quadratic operator.  Without actually solving the problem, we know that the eigenvalue spectrum is bounded below since the operator is Sturm---Liouville.  The same differential equation
would apply to the expansion about the negative action classical solution
except with the opposite sign on the operator.  Therefore its eigenvalue spectrum
is bounded above and the negative action solution is highly unstable.
We do not know how to handle this problem, which is another reason we do
not consider the negative action solution.

It can be easily shown that the quadratic operator about the positive action solution
has a zero eigenvalue and at least one negative eigenvalue.
The instability of the positive action classical solution due to a finite
number of negative eigenmodes is much preferable to the extremely unstable
negative action classical solution with an infinite number of negative
eigenmodes.  The dependence of the quasiclassical prefactor can
be evaluated if the eigenvalue spectrum is known in the first case.  The problem is undefined in the latter case.

We propose to avoid divergences and regularization prescriptions by only
considering contributions to the partition function from the classical solution
with positive action.  This is a prescription that we cannot completely
justify, but is supported by arguments due to Marolf \cite{Marolf}.
His arguments do not completely carry over to our case, because the Hamiltonian is not positive definite, but they lend credence to our assumption.

There is reason to believe that the apparent singularity of the Euclidean
partition function arises as a result of an incorrect continuation of the
partition function from Minkowski space.  
It has been shown \cite{Schleich1},\cite{MazurMottola} that the correct Euclidean
path integral resembles the conformally rotated naive
Euclidean path integral, except for an extra Faddeev-Popov determinant factor.  The path integral defined in this way is nonsingular and well defined.  Our prescription
leads to a well defined path integral and the classical solution agrees with the
conformally rotated path integral.

Finally, although we have discussed the first nonvanishing variations about the
classical solution in this minisuperspace model, we have not calculated this prefactor.
The general expansion about the De Sitter solution was studied in \cite{Gibbons&Perry} and the prefactor for our case could be determined using their results, but this is not required for our purposes.

\section{$\phi^4$ Theory}

A second, more general, minisuperspace model arises from the metric\cite{KlebanovEtal}:
\begin{equation}
ds^2=\left(\frac{4\pi G}{3}\phi^2 \delta_{ij}\right) dx^i \wedge dx^j.
\end{equation}
The Euclidean action in this case is:
\begin{equation}
\hat{I}=- \int d^4x [\frac{1}{2}(\partial\phi)^2 + \frac{1}{4}
g \phi^4]
\label{phi}
\end{equation}
where $g=-8\pi G \Lambda/9$.
The path integral is:
\begin{equation}
Z=\int D[\phi] \,\exp\left(\int d^4x [\frac{1}{2}(\partial\phi)^2 + \frac{1}{4} g \phi^4] \right).
\end{equation}

This is similar to the path integral for Euclidean $\phi^4$ theory except for
the unconventional sign.
The large order behavior of $\phi^4$ theory is known
\cite{Lipatov},\cite{Zinn-Justin}
to have factorial dependence on the order with alternating sign terms.  This
occurs because the theory is expanded about a complex saddle point\footnote{There
are no instantons in a real scalar field theory, but these calculations are based on the analytical continuation of such a field theory and so may
have complex instantons.}.  The path
integral in our case will retain the factorial dependence but, as in the
previous model, the alternating sign is absent.
The results are:
\begin{equation}
\label{2}
Z_K\propto\left(\frac{8\pi G \Lambda}{9} I(\phi_{cl})\right)^{-K} K!
\end{equation}
where $\phi_{cl}$ is the instanton:
\begin{equation}
\phi_{cl}(r)={\sqrt{\frac{9}{\pi G \Lambda}}}~~\frac{\lambda}{[1+\lambda^2 (r-r_0)^2]}
\end{equation}
with action $I(\phi_{cl})=\frac{3\pi}{\Lambda G}$, corresponding to Euclidean De
Sitter space.  This model intersects the previous model at its classical
solutions.

The quasiclassical prefactor can be determined from the results of \cite{Gibbons&Perry},\cite{Lipatov}.

\section{Conclusion and Wild Speculations}

The main result of this paper is the factorial dependence on order of higher order terms in the quasiclassical expansion of Euclidean gravity.  We have shown this in two minisuperspace models using only the value of the classical action.  This
dependence is characteristic of an asymptotic series.  This is just the type
of behavior we would expect of a low energy effective field theory.  The
factorial growth of expansion coefficients was assumed in \cite{Brandenberger&Zhitnitsky} to deal with problems of inflation.

The expansion may or may not be Borel summable depending on the classical
solution one expands about.  We chose to expand about the positive action
solution.  We assumed following \cite{Marolf} that only the positive action
solution contributes to the Euclidean path integral and expanded about that
solution,  but this assumption does not affect the general statement above.

In the first minisuperspace model we argued that at least one negative eigenmode exists, but left the calculation of quasiclassical prefactor for
the future.  The second minisuperspace model
is very similar to $\phi^4$ theory for which the first quasiclassical
prefactor can be obtained from \cite{Lipatov}.

The nonrenormalizable nature of gravity --- the most profound field
theoretic feature --- does not show up at the classical level
in the minisuperspace models considered in this paper (\ref{1},\ref{2}).  Indeed,
any ultraviolet (UV) divergences will appear in our
approach only at the level of calculation of
the one-loop quantum determinant. The corresponding divergences can be absorbed
in the standard way by redefining the original parameters of the
$\phi^4$ theory. It can be done explicitly
due to the renormalizability
of the obtained $\phi^4$ theory (\ref{phi}). The procedure is well defined,
but seems to involve an exchange of
limit between large cut-off and large order. We make a standard
field theory assumption that such an exchange is justified
and does not affect the main result of factorial growth. Note that a similar
assumption is not required for the first minisuperspace model which is reduced to a quantum mechanical system (\ref{qm}) rather than a field theory.

In general, however,
because of the dimensionality of the coupling constant $G$, one loop
calculations generate operators proportional not to the original action $R$, but  higher order terms such as $R^2,~ R_{\mu\nu}R^{\mu\nu}$. This is an explicit
manifestation of the non-renormalizability of gravity.
However, when gravity is treated as an effective field theory
 it can be renormalized
at any given order by absorbing the divergences into renormalized
values of the coefficients in the most general effective Lagrangian.
In this respect the theory resembles the effective chiral Lagrangian approach
where one can show that coefficients also exhibit
factorial growth\cite{Zhitnitsky}.  More importantly,
this growth is not affected by the UV divergences at least at the one loop level.

Indeed, in the semiclassical approximation the UV divergences appear only at the level of the calculation of the quantum determinant.
As usual, this determinant is divergent but
factorized from the classical part.  Therefore, the quantum part
is independent of the $k!$ behavior, which originates from the classical
action.
Technically, the factor related to ultraviolet regularization
at one-loop level will appear
in front of $k!$:
\begin{equation}
\label{d}
Z_k \, \sim\frac{1}{\epsilon}k! , \,\,\,\,\, \epsilon=d-4 .
\end{equation}
One can absorb this divergence at one loop level
by redefining any counterterm from the action. After that, formula
(\ref{d}) gives a finite result for arbitrary high terms and explicitly
demonstrates the expected $k!$ behavior.

One could stop here if we accept the Hawking viewpoint \cite{Hawking1}
that the dominant contribution to the functional integral can be represented
as a sum of background and one-loop terms {\bf only}! This point is motivated by
an idea that all classical solutions (i.e. all metrics) with all possible topologies
are dense in some sense in the space of all metrics. If this is the case,
one could then hope  (see \cite{Hawking1}) to pick out some
finite number of solutions which give the dominant contribution
to the path integral (spacetime foam picture).

This prescription for handling the nonrenormalizability of quantum gravity
has survived for many years.  Nevertheless, we believe it might be interesting to discuss
another possibility for treating this problem which should be considered,
at the moment, as wild speculation at best.

The basic idea, as before, is the observation
that the perturbation series (\ref{k}) is an asymptotic one.
Therefore, we can use
some integral  representation formula for this
      series.  For illustration purposes we
assume that the series is Borel summable and
we use a Borel representation for it \footnote{
We already mentioned that
the series (\ref{k}) which represents our system
most likely is not Borel summable.  We believe, however, that the
Borel non-summability of an expansion does not signal
an inconsistency or ambiguity of the theory.
The Borel prescription is just one of many summation methods
and need not be applicable everywhere.
Thus, some prescription, based on the physical considerations,
should be given
in order to evaluate an integral like this.  Some new physics
usually accompanies such a phenomenon, but we do not go
into details here.
Rather, we would like to
  mention the non-Borel summable example of the principal
chiral field theory at large N\cite{Fateev}.  In this case,
the explicit solution is known.  The coefficients
grow factorially with the order and the series is not Borel summable.
Nevertheless, the physical observables are perfectly well defined and
the exact result can be recovered by a special prescription which
uses a non-trivial procedure of analytic continuation.}:
\be
\label{Borel}
Z(g)=\sum_{K=0}^\infty Z_K g^K\sim
 \int_{0}^{\infty}\frac{f(t)dt}{t(t+g)}
\exp(-\frac{1}{t}),
\ee
where  a  function $f(t)$ is defined by moments
\be
Z_K\sim (-)^K\int_{0}^{\infty}\frac{f(t)dt}
{t^{K+1}}\exp(-\frac{1}{t})\sim (-)^K K!
\ee
and should be mild enough to preserve the main asymptotical behavior $\sim K!$. This would be
 the end of the story if we were to discuss a quantum mechanical problem.
However, we wish to discuss a nonrenormalizable field theory where, at the one-loop level, an UV divergence will appear in front
of the $K!$ factor, as in (\ref{d}).
However, the most singular behavior in gravity which could occur in front of
$G^{K}$ is not the one-loop divergence, $\frac{1}{\epsilon}$, we discussed previously, but rather the K-th loop divergence proportional
to $\frac{1}{\epsilon^K}$. Therefore, in general, we expect
the following structure for the K-th loop term  in quantum gravity:
\be
\label{e}
Z_K \, \sim\frac{K!}{\epsilon^K}c_0^{(K)}(1+c_1^{(K)}
\epsilon+ c_2^{(K)}\epsilon^2+...)
,~~~~   \epsilon=d-4 .
\ee
Now, if we use the Borel prescription (\ref{Borel}) to sum up this series,
then we would get a result of zero for this series:
 \be
\label{Borel1}
Z(g)=\sum_{K=0}^\infty Z_K \frac{G^K}{\epsilon^K}\sim
 \int_{0}^{\infty}\frac{\exp(-\frac{1}{t})
f(t)dt}{t(t+\frac{G}{\epsilon}(1+0(\epsilon))}
\sim\epsilon\rightarrow 0,
\ee
in spite of
the fact that each term on the left hand side diverges in the limit
$\epsilon=d-4\rightarrow 0$ and irrespective of the precise behavior of the
coefficients, $c_0^{(K)}$, which presumably can be modeled by a function $f(t)$.
The finite terms at each level apparently have very different analytical
structure (see \cite{Donoghue}) and should be treated separately
from UV divergent terms (\ref{Borel1}).

We are not pretending to have made a reliable analysis
of UV divergences in gravity in this letter.  Rather, we wanted to
point out that an asymptotic nature of the expansion might provide a
natural way to handle the problem of UV divergences in gravity.
As we mentioned, there is a possibility that each term in the series
is divergent, but the series itself is a well defined function.
At least, we cannot rule out this possibility from the very
beginning and we believe it deserves further investigation.

\end{document}